\documentclass[12pt]{article}
\usepackage{epsfig}
\begin{document}
\begin{center}
{\Large {\bf Holst action and Dynamical Electroweak symmetry breaking}


{
\vspace{1cm}
{ M.A.~Zubkov }\\
\vspace{.5cm} {\it  ITEP, B.Cheremushkinskaya 25, Moscow, 117259, Russia } }}
\end{center}

\begin{abstract}
We consider Poincare gravity coupled in a nonminimal way to spinors. The
gravitational action is considered that contains both Palatini and Holst terms.
Due to torsion the effective four - fermion interactions appear that may lead
to the left - right asymmetry and the condensation of fermions.  When the mass
parameter entering the mentioned terms of the gravitational action is at a Tev
scale the given construction may provide the dynamical Electroweak symmetry
breaking. This is achieved via an arrangement of all Standard Model fermions in
the left - handed Dirac spinors while the right - handed spinors are reserved
for the technifermions. Due to the gravitational action the technifermions are
condensed and, therefore, cause the appearance of gauge boson masses.
\end{abstract}

\section{Introduction}

Holst action \cite{Hehl,Holst:1995pc,imir,Rovelli,Khriplovich:2005jh} breaks
parity. Therefore, when Poincare gravity that contains Holst term in the action
is coupled to fermions, it may provide the appearance of parity breaking four -
fermion interactions
\cite{Freidel:2005sn,Randono:2005up,Mercuri:2006um,Alexandrov}. The four -
fermion interactions induced by Holst action were considered in
\cite{Xue,Alexander1,Alexander2} as a source of fermion condensation used,
mainly, in a cosmological background. In our previous paper \cite{Z2010_3} we
have suggested that nonminimal coupling of fermion fields to torsion
\cite{Shapiro} may provide condensation of the additional fermions (called
technifermions in an analogy with Technicolor theory (TC)
\cite{Technicolor,Technicolor_,Technicolor__}) and provide Dynamical
Electroweak Symmetry Breaking (DEWSB) if torsion mass parameter is at a Tev
scale.

Namely, we arrange all SM fermions in the left - handed components of the Dirac
spinors while the additional fermions (called technifermions) are arranged in
right - handed components of the spinors. If the parity breaking is admitted in
the torsion action, under natural assumptions this action has the form that
leads to appearance of the  asymmetry between the left-handed and the
right-handed fermions. Due to this asymmetry the four fermion interactions
provide condensation of the technifermions while do not affect qualitatively
the dynamics of the SM particles. As a result the Electroweak symmetry is
broken.

In the suggested approach the problems specific for TC, Extended Technicolor
(ETC)
\cite{ExtendedTechnicolor,ExtendedTechnicolor_,ExtendedTechnicolor__,ExtendedTechnicolor___,ExtendedTechnicolor____,walking,minimal_walking,minimal_walking_,minimal_walking__},
and Bosonic Technicolor \cite{Kagan,Dobrescu_Kagan,Z2010_2} are avoided.
However, the suggested approach contains its own difficulties. First, the
source of the parity violating action for torsion was not suggested. Second,
the effective theory that contains four - fermion interactions is to be treated
in a way similar to the effective Nambu - Jona - Lasinio model (NJL) of chiral
symmetry breaking \cite{ENJL} in TC. Namely, the NJL model is a
nonrenormalizable finite cutoff theory. And physical results depend on the
value of the cutoff $\Lambda_{\chi}$ that becomes an additional physical
parameter.

In the present paper we suggest a possible origin of the parity violating
action. Namely, in Poincare gravity the torsion field is to manifest itself
through the curvature of Riemann - Cartan space. There are only two possible
terms constructed of curvature that contain squared torsion. These two terms
are Palatini and Holst terms. We consider both these terms. In addition we
consider the nonminimal coupling of spinors to gravity that itself admits
parity violation. We consider several limiting cases and derive the conditions
under which the resulting four - fermion interactions are repulsive for SM
fermions and attractive for the technifermions (or the induced interactions for
SM fermions are negligible compared to that of the technifermions). In all
these cases the theory admits condensation of technifermions while SM fermions
remain massless.
 That's why we partially
resolve the first difficulty mentioned above. As for the second difficulty, we
cannot resolve it at the present moment and consider it as a subject of future
investigations.

The paper is organized as follows. In the 2-nd section we consider fermion
fields in Riemann-Cartan space.  In the 3-rd section we consider Holst action
and derive the four - fermion interactions that appear after integration over
torsion. In the $4$ - th section we consider several limiting cases and point
out the regions in space of couplings, where right-handed fermions may be
condensed while left - handed fermions remain massless. In the $5$ - th section
we introduce two kinds of spinors nonminimally coupled to gravity. The left -
handed components of these spinors are used to arrange both left - handed and
right - handed SM fermions while the right - handed components of these spinors
are used to arrange technifermions. We demonstrate how the resulting four -
fermion action can be written in terms of $4$ - component SM fermions and
technifermions. In the $6$ - th section we apply NJL technique to the four -
fermion interactions of technifermions and describe how chiral symmetry
breaking occurs. In section $7$ we introduce mass term for original spinors
that contain SM fermions as their left - handed components. We then derive the
mass term for the SM fermions. In section $8$ we end with the conclusions.

\section{Fermions in Riemann-Cartan space}

We consider the action of a massless Dirac spinor in Riemann - Cartan space in
the form \cite{Alexandrov}:

\begin{eqnarray}
S_f & = & \frac{i}{2}\int E \{ \bar{\psi} \gamma^{\mu} (\zeta - i\chi\gamma^5)
D_{\mu} \psi - [D_{\mu}\bar{\psi}](\bar{\zeta} - i\bar{\chi}\gamma^5)
\gamma^{\mu}\psi \} d^4 x \label{Sf}
\end{eqnarray}

Here $\zeta=\eta + i\theta$ and $\chi = \rho+i\tau$ are the coupling constants,
$E = {\rm det} E^a_{\mu}$, $E^a_{\mu}$ is the inverse vierbein, $\gamma^{\mu} =
E^{\mu}_a \gamma^a$, the covariant derivative is $D_{\mu} =
\partial_{\mu} + \frac{1}{4}(\omega_{\mu}^{ab}+C_{\mu}^{ab})\gamma_{[a}\gamma_{b]}$;
$\gamma_{[a}\gamma_{b]} =
\frac{1}{2}(\gamma_{a}\gamma_{b}-\gamma_{b}\gamma_{a})$. The torsion free spin
connection is denoted by $\omega_{\mu} $ while $C_{\mu}$ is the contorsion
tensor. They are related to $E^a_{\mu}$, Affine connection $\Gamma^{i}_{jk}$,
and torsion $T^a_{.\mu \nu}= T^{\rho}_{.\mu\nu}E^a_{\rho}$ as follows:
\begin{eqnarray}
\nabla_{\nu} E_{\mu}^{a} &=& \partial_{\nu}E^a_{\mu} - \Gamma^{\rho}_{\mu
\nu}E^a_{\rho} + \omega^{a}_{ . b\nu}E^b_{\mu}+ C^{a}_{ . b\nu}E^b_{\mu}=0\nonumber\\
\tilde{D}_{[\nu} E_{\mu]}^{a} &=& \partial_{[\nu}E^a_{\mu]} +
\omega^{a}_{.b[\nu }E^b_{\mu]}=0\nonumber\\
T^a_{.\mu \nu} & = & {D}_{[\nu} E_{\mu]}^{a} = \partial_{[\nu}E^a_{\mu]} +
\omega^{a}_{.b[\nu }E^b_{\mu]}+ C^{a}_{ . b[\nu}E^b_{\mu]}=C^{a}_{ .
b[\nu}E^b_{\mu]}
\end{eqnarray}

This results in:
\begin{eqnarray}
\Gamma^{\rho}_{\mu \nu}&=& \{^{\rho}_{\mu \nu}\} + C^{\rho}_{.\mu \nu}
\nonumber\\
C^{\rho}_{.\mu \nu} & = & \frac{1}{2}(T^{\rho}_{.\mu \nu}-T^{.\rho}_{\nu
.\mu}+T^{..\rho}_{\mu \nu})\nonumber\\
\{^{\alpha}_{\beta \gamma}\} & = &
\frac{1}{2}g^{\alpha\lambda}(\partial_{\beta}g_{\lambda
\gamma}+\partial_{\gamma}g_{\lambda \beta}-\partial_{\lambda}g_{\beta \gamma})\nonumber\\
\omega_{ab\mu} & = &\frac{1}{2}( c_{abc}-c_{cab}+c_{bca})E^{c}_{\mu}\nonumber\\
C_{ab\mu} & = &\frac{1}{2}( T_{abc}-T_{cab}+T_{bca})E^{c}_{\mu}
\end{eqnarray}
Here $c_{abc} = \eta_{ad}E^{\mu}_b E_c^{\nu}\partial_{[\nu}E^d_{\mu]}$;
$T_{abc} = \eta_{ad}E^{\mu}_b E_c^{\nu}T_{.\mu\nu}^d$; $g_{\mu \nu} =
E^a_{\mu}E^b_{\nu}\eta_{ab}$; $\Gamma^{\rho}_{\mu \nu}-\Gamma^{\rho}_{\nu \mu}
= T^{\rho}_{.\mu \nu}$; indices are lowered and lifted via $g$ and $E$ as
usual.

(\ref{Sf}) can be rewritten as follows:

\begin{eqnarray}
S_f & = & \frac{1}{2}\int E\{i\bar{\psi} \gamma^{\mu}(\zeta - i\chi\gamma^5)
\tilde{D}_{\mu} \psi - i[\tilde{D}_{\mu}\bar{\psi}](\bar{\zeta} -
i\bar{\chi}\gamma^5) \gamma^{\mu}\psi\nonumber\\&& + \frac{i}{4}
C_{abc}\bar{\psi}[\{ \gamma^{[a}\gamma^{b]},\gamma^c\}(\eta+\tau\gamma^5) - i[
\gamma^{[a}\gamma^{b]},\gamma^c](\theta-\rho\gamma^5)] \psi\} d^4 x \label{Sf1}
\end{eqnarray}
Here $\tilde{D}$ is the covariant derivative of general relativity. Next, we
obtain:
\begin{eqnarray}
S_f & = & \frac{1}{2}\int E\{i\bar{\psi} \gamma^{\mu}({\zeta} -
i{\chi}\gamma^5) \tilde{D}_{\mu} \psi -
i[\tilde{D}_{\mu}\bar{\psi}](\bar{\zeta} - i\bar{\chi}\gamma^5)
\gamma^{\mu}\psi\nonumber\\&& - \frac{1}{4}
C_{abc}\bar{\psi}[-2\epsilon^{abcd}\gamma^5 \gamma_d(\eta+\tau\gamma^5) +4
\eta^{c[a} \gamma^{b]}(\theta-\rho\gamma^5)] \psi\} d^4 x \label{Sf2}
\end{eqnarray}

Now let us introduce the irreducible components of torsion:
\begin{eqnarray}
S^i& =& \epsilon^{jkli}T_{jkl}\nonumber\\
T_i& =& T^j_{.ij}\nonumber\\
T_{ijk}& =& \frac{1}{3}(T_j \eta_{ik}-T_k\eta_{ij}) -
\frac{1}{6}\epsilon_{ijkl}S^l + q_{ijk}
\end{eqnarray}

In terms of $S$ and $T$ (\ref{Sf2}) can be rewritten as:
\begin{eqnarray}
S_f & = & \frac{1}{2}\int E\{i\bar{\psi} \gamma^{\mu}(\zeta - i\chi\gamma^5)
\tilde{D}_{\mu} \psi - i[\tilde{D}_{\mu}\bar{\psi}](\bar{\zeta} -
i\bar{\chi}\gamma^5) \gamma^{\mu}\psi\nonumber\\&& +
\frac{1}{4}\bar{\psi}[\gamma^5 \gamma_d (\eta S^d - 4\rho T^d) -(\tau
S^b+4\theta T^b) \gamma_{b}] \psi\} d^4 x \label{Sf22}
\end{eqnarray}

\section{Holst action and Dirac fermions}

Let us consider the Holst action:
\begin{equation}
S_T = -M_T^2 \int E E_a^{\mu}E_b^{\nu}G^{ab}_{\mu\nu} d^4x -
\frac{M_T^2}{\gamma}\int E E_a^{\mu}E_b^{\nu} {^*}G^{ab}_{\mu\nu}d^4x\label{ST}
\end{equation}

Here $G^{ab}_{\mu\nu} = [D_{\mu},D_{\nu}]$ is the $SO(3,1)$ curvature of
Riemann-Cartan space while ${^*}G^{ab}_{\mu\nu} = \frac{1}{2}
\epsilon^{ab}_{..cd}G^{cd}_{\mu\nu}$ is its dual tensor. In the absence of
torsion the second term is the integral of a total derivative and, therefore,
disappears from classical consideration. However, in presence of torsion, it
gives nontrivial part to the fermion interactions as will be seen later.

Now let us represent Holst action in terms of torsion and Riemannian curvature
 \cite{Mercuri:2006um}:

\begin{eqnarray}
S_T = M_T^2 \int E\{- R + \frac{2}{3}T^2 - \frac{1}{24}S^2  +
\frac{1}{3\gamma}TS \}d^4x + \tilde{S}
\end{eqnarray}
Here $R$ is Riemannian scalar curvature, $\tilde{S}$ contains the terms that
depend on $q$ and the so-called Nieh - Yan invariant.

Let us now suppose for a moment that there are no other terms that depend on
torsion in the gravitational action. Then integration over torsion degrees of
freedom can be performed for the system that consists of the Dirac fermion
coupled to gravity. The result of this integration is:

\begin{eqnarray}
S_f& = & \frac{1}{2}\int E\{i\bar{\psi} \gamma^{\mu}(\zeta - i\chi\gamma^5)
\tilde{D}_{\mu} \psi - i[\tilde{D}_{\mu}\bar{\psi}](\bar{\zeta} -
i\bar{\chi}\gamma^5) \gamma^{\mu}\psi\}d^4x\nonumber\\&& -
\frac{3\gamma^2}{(1+\gamma^2)32M_T^2} \int E
\{V^2[\theta^2-\tau^2+\frac{2\theta\tau}{\gamma}] +
A^2[\rho^2-\eta^2-\frac{2\eta\rho}{\gamma}]\nonumber\\&&+2AV[\theta\rho+\tau\eta+\frac{\rho\tau-\theta\eta}{\gamma}]
\} d^4x + S_{eff}[E]\label{F42}
\end{eqnarray}
Here we have defined:
\begin{eqnarray}
V_{\mu} & = & \bar{\psi} \gamma_{\mu}  \psi \nonumber\\A_{\mu} & = &
\bar{\psi}\gamma^5 \gamma_{\mu}  \psi
\end{eqnarray}
$S_{eff}$ is the effective action that depends on metric field only. It comes
from the functional determinant after the integration over torsion is
performed. If terms that contain derivatives of torsion are absent, $S_{eff}$
is reduced to the renormalization of the cosmological constant. For this reason
we omit it below. The four fermion term of (\ref{F4}) differs from that of
obtained in \cite{Alexandrov} by the overall sign and the sign of the immirzi
parameter $\gamma$ due to the difference in the definition of  action
(\ref{ST}).

Now let us introduce the right-handed and the left-handed currents:
\begin{eqnarray}
J_+^{\mu} & = & \bar{\psi}_+ \gamma^{\mu}  \psi_+ \nonumber\\J_-^{\mu} & = &
\bar{\psi}_- \gamma^{\mu}  \psi_-
\end{eqnarray}
Here $\psi_+$ is the right - handed component of $\psi$ while $\psi_-$ is the
left - handed component. In the further consideration we consider the case
$E^a_{\mu}=\delta^a_{\mu}$, and $\omega_{\mu}=0$. We also rescale left - handed
and right - handed components of $\psi$:
\begin{equation}
\psi_-\rightarrow \frac{1}{\sqrt{\eta+\tau}}\psi_-\, ; \,\psi_+\rightarrow
\frac{1}{\sqrt{\eta-\tau}}\psi_+
\end{equation}

Now (\ref{F42}) can be rewritten as follows:
\begin{eqnarray}
S_f& = & \frac{1}{2}\int \{i\bar{\psi} \gamma^{\mu} \partial_{\mu} \psi -
i[\partial_{\mu}\bar{\psi}] \gamma^{\mu}\psi\} d^4x\nonumber\\&& -
\frac{3\gamma^2}{(1+\gamma^2)32M_T^2} \int \{J_+^2 [-1 +
\frac{(\theta+\rho)^2}{(\eta-\tau)^2}
-\frac{2(\theta+\rho)}{(\eta-\tau)\gamma}] \nonumber\\&&+ J_-^2 [-1 +
\frac{(\theta-\rho)^2}{(\eta+\tau)^2}
+\frac{2(\theta-\rho)}{(\eta+\tau)\gamma}]\nonumber\\&&+ 2J_+J_-[1 +
\frac{\theta^2-\rho^2}{\eta^2-\tau^2}
+\frac{\theta+\rho}{(\eta-\tau)\gamma}-\frac{\theta-\rho}{(\eta+\tau)\gamma}]
\} d^4x \label{F4}
\end{eqnarray}

  The next step would be to consider the fermions coupled to Poincare gravity
 with higher derivative action for the gravitational fields. It is well - known
 that such a theory could be  renormalizable in the presence of terms quadratic
 in curvature \cite{Sesgin,Elizalde}. The models
 of this kind, however, suffer from the so-called unitarity problem. Moreover,
 the possibility to obtain Newtonian limit is questionable. Nevertheless, below
 we suppose that a self-consistent Poincare gravity theory exists, probably,
 with the elements involved that are not known at present. Our main supposition
 here is that Poincare gravity has two different scales. The first
 one (around Plamk mass $m_p$) is related to Riemannian geometry and produces Einstein - Hilbert action
 in the low energy approximation. The second scale $\Lambda_{\chi}$ is related to dynamical
 torsion theory. The effective charge entering the term of the action with the
 derivative of torsion depends on the ratio $\epsilon/\Lambda_{\chi}$, where $\epsilon$ is the energy scale of the considered physical process.
 As it will be explained further we expect that $\Lambda_{\chi}$ is at most one order of
 magnitude larger than $1$ Tev. In addition we also have the mass parameter  $M_T$
 of  (\ref{ST}) that is supposed to be at a Tev scale.

  At the scale $\Lambda_{\chi}$ in addition to (\ref{ST})
  the torsional part of the whole
 gravitational action contains terms that depend on the derivatives of $T$ and $S$.
  This means, in particular, that  the following terms may enter the action:
\begin{eqnarray}
S_T &=&  \beta_1 \int E G^{abcd}G_{abcd}d^4x+\beta_2 \int
EG^{abcd}G_{cdab}d^4x\nonumber\\&& +\beta_3 \int EG^{ab}G_{ab}d^4x+\beta_4 \int
EG^{ab}G_{ba}d^4x\nonumber\\&& +\beta_5 \int EG^2d^4x+\beta_6 \int
EA^{abcd}A_{abcd}d^4x\label{ST2}
\end{eqnarray}
with coupling constants $\beta_{1,2,3,4,5,6}$. Here
$G^{abcd}=E^c_{\mu}E^d_{\nu}G^{ab}_{\mu\nu}$, $G^{ac}=G^{abc}_{...b}$, $G =
G^a_a$, $A_{abcd} =
\frac{1}{6}(G_{abcd}+G_{acdb}+G_{adbc}+G_{bcad}+G_{bdca}+G_{cdab})$. Actually,
action (\ref{ST2}) is the most general quadratic in curvature action that does
not contain Parity breaking.

 Then the integration over torsion
 leads to (\ref{F4}) in the low energy limit $\epsilon << \Lambda_{\chi}$. That's why the obtained theory
 with the four - fermion action is only the effective low energy approximation
 that works at the energies much less than the scale $\Lambda_{\chi}$ of Poincare gravity.

\section{Limiting cases}

In this section we consider different limiting cases of (\ref{F4}). Our aim is
to find out the possibility that there exist attractive interactions between
the right - handed fermions while the interaction between the left-handed
fermions is either repulsive or is negligible compared to that of the right
-handed fermions. We also need the interaction between the right-handed and the
left-handed fermions is negligible. All this is needed in order to provide the
condensation of right-handed fermions used in the next sections in order to
provide DEWSB.

\subsection{Einstein - Cartan gravity}

This case corresponds to infinite immirzi parameter $\gamma$. We have:
\begin{eqnarray}
S_f& = & \frac{1}{2}\int \{i\bar{\psi} \gamma^{\mu} \partial_{\mu} \psi -
i[\partial_{\mu}\bar{\psi}] \gamma^{\mu}\psi\} d^4x\nonumber\\&& -
\frac{3}{64M_T^2} \int \{J_+^2 [-1 + \frac{(\theta+\rho)^2}{(\eta-\tau)^2} ]
\nonumber\\&&+ J_-^2 [-1 + \frac{(\theta-\rho)^2}{(\eta+\tau)^2}
]\nonumber\\&&+ 2J_+J_-[1 + \frac{\theta^2-\rho^2}{\eta^2-\tau^2} ] \} d^4x
\label{F4E}
\end{eqnarray}

In order for the cross term to vanish we need $1 +
\frac{\theta^2-\rho^2}{\eta^2-\tau^2} =0$ that is $|\zeta|=|\chi|$. We find
repulsive interactions between $J_-$ and attractive interactions between $J_+$
at
\begin{eqnarray}
&&|\zeta|=|\chi|\nonumber\\&& |{\rm Re} \zeta + {\rm Im} \chi|<|{\rm Im} \zeta
- {\rm Re} \chi|\nonumber\\&& |{\rm Re} \zeta - {\rm Im} \chi|>|{\rm Im} \zeta
+ {\rm Re} \chi|\label{C1}
\end{eqnarray}

Interactions between $J_-$ disappear if $|{\rm Re} \zeta + {\rm Im} \chi|=|{\rm
Im} \zeta - {\rm Re} \chi|$.

\subsection{Holst term in the action}
Let us consider the situation, when Palatini action is absent and only the
Holst term is present. We have:

\begin{eqnarray}
S_f& = & \frac{1}{2}\int \{i\bar{\psi} \gamma^{\mu} \partial_{\mu} \psi -
i[\partial_{\mu}\bar{\psi}] \gamma^{\mu}\psi\} d^4x\nonumber\\&& -
\frac{3\gamma}{32M_T^2} \int \{J_+^2 [-\frac{2(\theta+\rho)}{(\eta-\tau)}] +
J_-^2 [\frac{2(\theta-\rho)}{(\eta+\tau)}]\nonumber\\&&+
2J_+J_-[\frac{\theta+\rho}{(\eta-\tau)}-\frac{\theta-\rho}{(\eta+\tau)}] \}
d^4x \label{F4H}
\end{eqnarray}

In order for the cross term to vanish we need
\begin{equation}
\gamma\frac{\theta+\rho}{(\eta-\tau)}=\gamma\frac{\theta-\rho}{(\eta+\tau)}=\alpha
\label{C2}
\end{equation}
where $\alpha$ is the new coupling constant.

 Then we
have:
\begin{eqnarray}
S_f& = & \frac{1}{2}\int \{i\bar{\psi} \gamma^{\mu} \partial_{\mu} \psi -
i[\partial_{\mu}\bar{\psi}] \gamma^{\mu}\psi\} d^4x\nonumber\\&& -
\frac{3\alpha}{16M_T^2} \int \{J_-^2 - J_+^2
 \}
d^4x \label{F4H}
\end{eqnarray}

We find repulsive interactions between $J_-$ and attractive interactions
between $J_+$ at
\begin{eqnarray}
\gamma\frac{\theta+\rho}{(\eta-\tau)}=\gamma\frac{\theta-\rho}{(\eta+\tau)}=\alpha>0
\end{eqnarray}

\subsection{General case}

In general case in order to have  attractive interactions between
technifermions and repulsive interactions between SM fermions we need:
\begin{eqnarray}
&&\frac{\gamma^2}{(1+\gamma^2)2}\{1 - \frac{(\theta+\rho)^2}{(\eta-\tau)^2}
+\frac{2(\theta+\rho)}{(\eta-\tau)\gamma}\}=\alpha_+>0\nonumber\\&&\frac{\gamma^2}{(1+\gamma^2)2}
\{-1 + \frac{(\theta-\rho)^2}{(\eta+\tau)^2}
+\frac{2(\theta-\rho)}{(\eta+\tau)\gamma}\}=\alpha_->0\label{C3}
\end{eqnarray}

In order to exclude the cross term $J_+J_-$ we need
\begin{equation}
|\frac{\gamma^2}{(1+\gamma^2)2}\{1 + \frac{\theta^2-\rho^2}{\eta^2-\tau^2}
+\frac{\theta+\rho}{(\eta-\tau)\gamma}-\frac{\theta-\rho}{(\eta+\tau)\gamma}\}|=|\alpha_{+-}|<<
\alpha_+\label{req}
\end{equation}

 Now let us
consider the following region of couplings:
\begin{eqnarray}
&&|\eta-\tau|<<1\nonumber\\&&|\eta+\tau|\sim|\theta-\rho|\sim|\theta+\rho|\sim
1\nonumber\\&&\gamma\sim 1
\end{eqnarray}
In this domain we can neglect the terms with $J_-^2$ and $J_-J_+$, and the
action receives the form:
\begin{eqnarray}
S_f& = & \frac{1}{2}\int \{i\bar{\psi} \gamma^{\mu} \partial_{\mu} \psi -
i[\partial_{\mu}\bar{\psi}] \gamma^{\mu}\psi\} d^4x\nonumber\\&& -
\frac{3\gamma^2}{(1+\gamma^2)32M_T^2}\frac{(\theta+\rho)^2}{(\eta-\tau)^2} \int
J_+^2
 d^4x \label{F41}
\end{eqnarray}

This is a repulsive interaction between the right-handed fermions. In order to
obtain attractive interactions one may change the overall sign in (\ref{ST}).
Formally this is equivalent to the change $M_T \rightarrow i M_T$. It is worth
mentioning that this situation corresponds to the sign of Palatini action
opposite to the conventional one.

\subsection{Discussion}

If action (\ref{ST}) is present with finite $\gamma$, while the fermion action
contains nonzero bare constants $\zeta$ and $\chi$, all of the effective
coupling constants $\theta,\eta,\rho,\tau$ receive contributions from loop
corrections due to the dynamical torsion. Of course, this may destroy the
conditions (\ref{C1}), (\ref{C2}), (\ref{req}), (\ref{C3}) and a kind of fine
tuning is necessary to keep the precise (or, almost precise) requirement
(\ref{req}).

A particulary interesting case is when the Palatini action and the bare
coupling $\chi$ are absent. So, we have the only dimensionless coupling $\alpha
= \alpha_+ = \alpha_-= \frac{\theta\gamma}{\eta}$. The necessary condition to
be imposed on the terms of the action that depend on the derivatives of torsion
is that they do not produce the Palatini action in the low energy limit. This
condition can be fulfilled, in particular, if the given terms are conformal
invariant. It is worth mentioning here that the usual conformal gravity at a
first look contradicts with the Newtonian limit. However, this difficulty may
be overcomed, in principle (see, for example, \cite{ConformalGrav}). As for the
effective coupling $\chi$, it may appear due to loop corrections because the
torsional action breaks Parity. So, if this limiting case is chosen it is
necessary to consider renormalization group trajectories for $\chi$ in order to
find out the domain of the theory, where effective coupling $\chi(\epsilon)$
vanishes. Then we are left with the only requirement $\alpha
> 0$.

\section{Composite Dirac fields}

Below we assume that due to the gravitational action at the considered energies
 the translational connection $E^a_{\mu}$ is close to
$\delta^a_{\mu}$  while usual Christoffel symbols vanish. Let us  consider
 two Dirac spinors $\psi$ and $\phi$ coupled
to gravity.   Then we consider the fermion action of the form:
\begin{eqnarray}
S_f & = & \frac{1}{2}\int \{i\bar{\psi} \gamma^{\mu} ({\zeta} -
i{\chi}\gamma^5)D_{\mu} \psi - i[D_{\mu}\bar{\psi}](\bar{\zeta} -
i\bar{\chi}\gamma^5) \gamma^{\mu}\psi \} d^4 x \nonumber\\&& +\frac{1}{2}\int
\{i\bar{\phi^c} \gamma^{\mu} ({\zeta} - i{\chi}\gamma^5)D_{\mu} \phi^c -
i[D_{\mu}\bar{\phi^c}](\bar{\zeta} - i\bar{\chi}\gamma^5) \gamma^{\mu}\phi^c \}
d^4 x \label{Sf2}
\end{eqnarray}

Here $\phi^c = i \gamma^2  \left(\begin{array}{c}\phi_-\\
\phi_+\end{array}\right)^*=\left(\begin{array}{c}i\sigma^2 \phi^*_+\\
-i\sigma^2 \phi^*_-\end{array}\right)$. Below we use the following
representation of
$\gamma$ matrices: $\gamma^{\mu} = \left(\begin{array}{cc}0&\sigma^{\mu}\\
\bar{\sigma}^{\mu}&0\end{array}\right)$, where $\bar{\sigma}^0 = \sigma^0 = 1;
\bar{\sigma}^i = -\sigma^i \, (i=1,2,3)$; $\gamma^5 = -i\gamma^0\gamma^1\gamma^2\gamma^3=\left(\begin{array}{cc}1&0\\
0&-1\end{array}\right)$.

According to the previous sections integration over torsion after suitable
rescaling of fermion fields leads to
\begin{eqnarray}
&&S_f  =  \int \{i\psi_+^+ {\sigma}^{\mu} \partial_{\mu} \psi_+  +i\psi_-^+
\bar{\sigma}^{\mu}
\partial_{\mu} \psi_-
+ i\phi_+^+ {\sigma}^{\mu}
\partial_{\mu} \phi_+  +
i\phi_-^+ \bar{\sigma}^{\mu} \partial_{\mu} \phi_-  \nonumber\\&&
+\frac{3\alpha_+}{16M_T^2}({\psi}^+_+{\sigma}^i \psi_+-{\phi}^+_-\bar{\sigma}^i
\phi_-)^2-\frac{3\alpha_-}{16M_T^2}({\phi}^+_+{\sigma}^i
\phi_+-{\psi}^+_-\bar{\sigma}^i \psi_-)^2
 \} d^4 x
\label{Sf_2}
\end{eqnarray}

Now let us compose new spinors $\psi_t = \left(\begin{array}{c}\phi_-\\
\psi_+\end{array}\right)$ and $\psi_s = \gamma^5\left(\begin{array}{c}\psi_-\\
\phi_+\end{array}\right)$. Then we come to the following  expression for the
effective action:

\begin{eqnarray}
S_f & = & \int \{i\bar{\psi}_s \gamma^{\mu} \partial_{\mu} \psi_s -
\frac{3\alpha_-}{16M_T^2}(\bar{\psi}_s\gamma^i \gamma^5
\psi_s)(\bar{\psi}_s\gamma_i \gamma^5  \psi_s)\} d^4 x \nonumber\\&&+ \int
\{i\bar{\psi}_t \gamma^{\mu}
\partial_{\mu} \psi_t  +
\frac{3\alpha_+}{16M_T^2}(\bar{\psi}_t\gamma^i \gamma^5
\psi_t)(\bar{\psi}_t\gamma_i \gamma^5  \psi_t)\} d^4 x \label{Sf22}
\end{eqnarray}

As in the previous section we assume  $\alpha_-,\alpha_+>0$. Then the
four-fermion interaction for the field $\psi_t$ is attractive while the
interaction terms for $\psi_s$ are repulsive. This opens a possibility that
$\psi_t$ is condensed while $\psi_s$ is not condensed.

\section{Electroweak symmetry breaking }

Let us arrange all left - handed fermions and right - handed fermions of the
Standard Model in the left - handed parts of Dirac spinors. Correspondingly,
the additional fields are arranged within the right-handed parts of the given
spinors. We call the mentioned additional fermion fields technifermions. The
effective low energy action has the form:
\begin{eqnarray}
&&S_f  =  \int \{i\bar{\psi}^a_s \gamma^{\mu} D_{\mu} \psi^a_s -
\frac{3\alpha_-}{16M_T^2}(\bar{\psi}^a_s\gamma^i \gamma^5
\psi^a_s)(\bar{\psi}^b_s\gamma_i \gamma^5  \psi^b_s)\} d^4 x \nonumber\\&&
+\int \{i\bar{\psi}^a_t \gamma^{\mu} D_{\mu} \psi^a_t  +
\frac{3\alpha_+}{16M_T^2}(\bar{\psi}^a_t\gamma^i \gamma^5
\psi^a_t)(\bar{\psi}^b_t\gamma_i \gamma^5  \psi^b_t)\} d^4 x\label{Sf22s}
\end{eqnarray}
Here indices $a,b$ enumerate the mentioned Dirac spinors while the derivative
$D$ contains all Standard Model gauge fields. Let us
 apply Fierz transformation to the four fermion term of (\ref{Sf22s}):
\begin{eqnarray}
S_{4}  &=&  \int \{ -\frac{3\alpha_-}{16M_T^2}(\bar{\psi}^a_s\gamma^i \gamma^5
\psi^a_s)(\bar{\psi}^b_s\gamma_i \gamma^5  \psi^b_s)\} d^4 x \nonumber\\&&
+\int \{ \frac{3\alpha_+}{16M_T^2}(\bar{\psi}^a_t\gamma^i \gamma^5
\psi^a_t)(\bar{\psi}^b_t\gamma_i \gamma^5  \psi^b_t)\} d^4 x\nonumber\\
&=& \frac{3\alpha_+}{16M_T^2}\int\{4
(\bar{\psi}^a_{t,L}\psi^b_{t,R})(\bar{\psi}^b_{t,R} \psi^a_{t,L})\nonumber\\&&
+[(\bar{\psi}^a_{t,L}\gamma_i\psi^b_{t,L})(\bar{\psi}^b_{t,L}\gamma^i
\psi^a_{t,L})+(L\leftarrow \rightarrow R)]\} d^4 x\nonumber\\&&
-\frac{3\alpha_-}{16M_T^2}\int\{4(\bar{\psi}^a_{s,L}\psi^b_{s,R})(\bar{\psi}^b_{s,R}
\psi^a_{s,L})\nonumber\\&&
+[(\bar{\psi}^a_{s,L}\gamma_i\psi^b_{s,L})(\bar{\psi}^b_{s,L}\gamma^i
\psi^a_{s,L})+(L\leftarrow \rightarrow R)]\} d^4 x
\end{eqnarray}
In this form the action has  the form similar to the extended NJL model for
$\psi_t$ (see Eq. (4), Eq. (5), Eq. (6) of \cite{ENJL}) (with negative $G_V$,
though). In addition we have the repulsive interactions between $\psi_s$. In
the absence of the Standard Model (SM) gauge fields the $SU({\cal N})_L\otimes
SU({\cal N})_R$ symmetry of (\ref{Sf22s}) is broken down to $SU({\cal N})_V$
(here ${\cal N}$ is the total number of SM fermions). The SM interactions act
as a perturbation.

 For the purpose of the further consideration we
denote by $N_t = 24$ the number of technifermions; $G_S = \frac{3\alpha_+
N_t\Lambda^2_{\chi}}{16M_T^2\pi^2}$; $G_V =-\frac{1}{4}G_S$. Here
$\Lambda_{\chi}$ is the cutoff that is now the physical parameter of the model.
Its value depends on the details of physics that provides the appearance of the
four - fermion interactions. In our case $\Lambda_{\chi}$ is to be calculated
within the (unknown) Poincare gravity theory. We also denote $g_s =
\frac{4\pi^2G_S}{N_t\Lambda_{\chi}^2}=\frac{3\alpha_+}{4M_T^2}$.

Next, the auxiliary fields $H$, $L_i$, and $R_i$ are introduced and the new
action for $\psi_t$ has the form:
\begin{eqnarray}
S_{4,t}  &=& \int\{ -(\bar{\psi}^a_{t, L}H^+_{ab} \psi^b_{R} + (h.c.)) -
\frac{4 M_T^2}{3\alpha_+} \, H_{ab}^+H_{ab}\}d^4x \nonumber\\&&
+\int\{(\bar{\psi}^a_{t,L}\gamma^i L^{ab}_i\psi^b_{t,L})
-\frac{4M_T^2}{3\alpha_+} {\rm Tr}\,L^i L_{i} +(L\leftarrow \rightarrow R)\}
d^4 x\label{eff}
\end{eqnarray}

Integrating out fermion fields we arrive at the effective action for the
mentioned auxiliary fields (and the source currents for fermion bilinears). The
resulting effective action receives its minimum at $H = m_t {\bf 1}$, where
$m_t$ plays the role of the technifermion mass (equal for all technifermions).

We apply the following regularization:
\begin{equation}
\frac{1}{p^2+m^2} \rightarrow \int_{\frac{1}{\Lambda_{\chi}^2}}^{\infty} d\tau
e^{-\tau (p^2+m^2)}
\end{equation}

With this regularization the expression for the condensate of $\psi_t$  is
(after the Wick rotation):
\begin{eqnarray}
<\bar{\psi}_t\psi_t> &=& N_t \int\frac{d^4p}{(2\pi)^4}\frac{1}{p\gamma + m}=
-N_t m_t\int\frac{d^4p}{(2\pi)^4}\frac{1}{p^2 + m^2_t}\nonumber\\&=&-N_t
m_t\int_{\frac{1}{\Lambda_{\chi}^2}}^{\infty} d\tau
\int\frac{d^4p}{(2\pi)^4}e^{-\tau(p^2 + m^2_t)}\nonumber\\&=&
 -\frac{N_t}{16\pi^2}4m_t^3
\Gamma(-1,\frac{m_t^2}{\Lambda_{\chi}^2})\label{GM}
\end{eqnarray}

Here $\Gamma(n,x) = \int_x^{\infty}\frac{dz}{z}e^{-z}z^{n}$. The gap equation
is:
\begin{equation}
m_t = -g_s<\bar{\psi}_t\psi_t>
\end{equation}

That is
\begin{equation}
m_t = G_S m_t
\{\exp(-\frac{m_t^2}{\Lambda_{\chi}^2})-\frac{m_t^2}{\Lambda_{\chi}^2}
 \Gamma(0,\frac{m_t^2}{\Lambda_{\chi}^2})\}
\end{equation}
It does not depend on $G_V$. Obviously, there exists the critical value of
$G_S$: at $G_S > 1$ the gap equation has the nonzero solution for $m_t$ while
for $G_S < 1$ it has not. This means that in this approximation the
condensation of technifermions occurs at

\begin{equation}
M_T < M_T^{\rm critical} = \sqrt{3\alpha_+ N_t}\frac{\Lambda_{\chi}}{4\pi}\sim
\sqrt{\alpha_+} \Lambda_{\chi}\label{cond}
\end{equation}
For example, at $\Lambda_{\chi}\sim 10$ Tev and $\alpha_+\sim 1/100$ we may
have $M_T^{\rm critical} \sim 1 $ Tev.

The technipion decay constant $F_T$ in the given approximation is:
\begin{equation}
F_t = \frac{N_t m_t^2}{4\pi^2}
 \Gamma(0,\frac{m_t^2}{\Lambda_{\chi}^2})
\end{equation}
Therefore,
\begin{equation}
F^2_t =
\frac{N_t\Lambda_{\chi}^2}{4\pi^2}e^{-\frac{m^2_t(M_T,\Lambda_{\chi})}{\Lambda_{\chi}^2}}
- \frac{4 M_T^2}{3\alpha_+}
\end{equation}
In order to have appropriate values of $W$ and $Z$ - boson masses we need
$F_T\sim 250$ Gev. At $M_T = M_T^{\rm critical}$ we have $m_t = 0$ and $F_T =
0$. When $M_T$ is decreased, $m_t$ increases and reaches the value around
$\Lambda_{\chi}$ somewhere at $M_T = M_T^{\rm critical}/2$. At this point
$F_T\sim \frac{\sqrt{N_t}\Lambda_{\chi}}{4\pi}$. As $\Lambda_{\chi} > 1$ Tev we
need $\frac{ M_T^{\rm critical}-M_T }{M_T^{\rm critical}} << 1$. Usual
naturalness requirement here means that $x = \frac{ [M_T^{\rm
critical}]^2-M^2_T }{[M_T^{\rm critical}]^2} \sim 0.1$. Smaller values of this
ratio would be considered unnatural. At small $x$ we have: $F_T \sim
\sqrt{N_t/2} \frac{\Lambda_{\chi}}{2\pi}x \sim 0.25$ Tev. Thus naturalness
 forbids to consider extremely large values of $\Lambda_{\chi}$
(say, of the order of Plank mass). That's why we bound ourselves by the values
of $\Lambda_{\chi}$ between $1$ Tev and $10$ Tev.

Negative $G_V$ leads to the appearance of the term in the action with
$(\rho_L^2 + \rho_R^2)$, where $\rho^{ab}_L =
(\bar{\psi}^a_{t,L}\gamma^0\psi^b_{t,L})$ and $\rho^{ab}_R =
(\bar{\psi}^a_{R,L}\gamma^0\psi^b_{R,L})$ are the densities of right-handed and
left - handed technifermions. This is an attractive interaction that
qualitatively corresponds to a positive shift of the chemical potential $\mu$.
That's why negative $G_V$ moves chiral symmetry restoration to smaller values
of $\mu$. However, we expect this change at $G_V =-\frac{1}{4}G_S$ does not
affect physics at $\mu = 0$ although this is to be the subject of an additional
investigation.

In the absence of SM interactions the relative orientation of the SM gauge
group $G_W = SU(3)\otimes SU(2)\otimes U(1)$ and $SU({\cal N})_V$ from
$SU({\cal N})_L\otimes SU({\cal N})_R \rightarrow SU({\cal N})_V$ is
irrelevant. However, when the SM interactions are turned on, the effective
potential due to exchange by SM gauge bosons depends on this relative
orientation. Minimum of the potential is achieved in the true vacuum state and
defines the pattern of the breakdown of $G_W$. This process is known as vacuum
alignment (see, for example, \cite{Align, Align1}). The effective potential is
\cite{Align}:
\begin{eqnarray}
V(U) &=& 4 \sum_{\alpha = SU(3), SU(2), U(1); \, k} e_{\alpha}^2 \, {\rm Tr} \,
(\theta^{\alpha, k}_L U \theta^{\alpha, k}_R U^+) \,\nonumber\\&&
(-\frac{i}{2}) \int d^4 x \Delta^{\mu \nu} (x) <0|T[ J^A_{\mu L} J^A_{\nu R}
|0>\nonumber\\&& = -\frac{3}{32\pi^2} (F^2 \Delta^2)\sum_{\alpha = SU(3),
SU(2), U(1); \, k}e_{\alpha}^2 \, {\rm Tr} \, (\theta^{\alpha, k}_L U
\theta^{\alpha, k}_R U^+)
\end{eqnarray}
There is no sum over $A$ here. $\theta^{\alpha, k}_{L,R}$ are generators of
$G_W$, $\Delta^{\mu \nu} (x)$ is the gauge boson propagator, $J^A_{\mu
L;R}=(\bar{\psi}^a_{t,L;R}\lambda_{ab}^A\gamma_i\psi^b_{t,L;R})$ are
technifermion currents; matrices $\lambda_{ab}^A$ are generators of  $ SU({\cal
N})$. $U \in SU({\cal N})$ defines relative orientation of $ SU({\cal N})_V$
and $G_W$. $F$ - is the technipion constant. In general case $\Delta^2$  may be
negative. However, in \cite{Align} arguments are given in favor of
$\Delta^2>0$. Namely, it was shown that if the technicolor interactions are
renormalizable and asymptotic free, then the spectral function sum rules take
place. Then under assumption that in the spectral functions correspondent to
vector and axial vector channels of $<0|T[ J^A_{\mu L} J^A_{\nu R} |0>$ single
intermediate states dominate, one finds $\Delta^2>0$. In our case dynamical
torsion  plays the role of the technicolor interactions. That's why we need
some suppositions about the dynamical torsion theory. In particular, if we
require that this theory is renormalizable and asymptotic free (as it should in
order to be self - consistent) and that two intermediate states dominate in the
mentioned above correlator, we also have $\Delta^2>0$. Under this supposition
in a way similar to that of \cite{Align} we come to the conclusion that $G_W$
is broken in a minimal way. This means that the subgroups of $G_W$ are not
broken unless they should. The form of the condensate requires that $SU(2)$ and
$U(1)$ subgroups are broken. That's why in a complete analogy with $SU(N_{TC})$
Farhi - Susskind model Electroweak group in our case is broken correctly while
$SU(3)$ group remains unbroken.

\section{Mass term }

In this section we consider the possibility to give masses to the SM fermions.
Namely, let us consider the action with an additional mass term for spinors
$\psi$ and $\phi$:

\begin{eqnarray}
S_f & = & \int \{\frac{i}{2}\bar{\psi}_a \gamma^{\mu}({\zeta} -
i{\chi}\gamma^5) D_{\mu} \psi_a + (c.c.) \} d^4 x \nonumber\\&& +\int
\{\frac{i}{2}\bar{\phi^c}_b \gamma^{\mu}({\zeta} - i{\chi}\gamma^5)
\bar{D}_{\mu} \phi_b^c + (c.c.)\} d^4 x \nonumber\\&& - \int(\delta_{a
a^{\prime}}\bar{\psi}_{a} \psi_{a^{\prime}} + {\bf f}_{b
b^{\prime}}\bar{\phi}_{b} \phi_{b^{\prime}}) m_0 d^4 x \label{Sf2SM__}
\end{eqnarray}
Here $m_0$ is the constant of the dimension of mass while ${\bf f}$ is the
hermitian matrix of couplings. Integrating over torsion we obtain:
\begin{eqnarray}
&&S_f  =  \int \{i\bar{\psi}^a_s \gamma^{\mu} D_{\mu} \psi^a_s -
\frac{3\alpha_-}{16M_T^2}(\bar{\psi}^a_s\gamma^i \gamma^5
\psi^a_s)(\bar{\psi}^b_s\gamma_i \gamma^5  \psi^b_s)\} d^4 x \nonumber\\&&+
\int \{i\bar{\psi}^a_t \gamma^{\mu} D_{\mu} \psi^a_t  +
\frac{3\alpha_+}{16M_T^2}(\bar{\psi}^a_t\gamma^i \gamma^5
\psi^a_t)(\bar{\psi}^b_t\gamma_i \gamma^5  \psi^b_t)\} d^4 x \nonumber\\&& -
\frac{m_0}{\sqrt{[{\rm Re}\zeta]^2-[{\rm Im} \chi]^2}}\int(\bar{\psi}_{s,a}
[\frac{\delta_{a a^{\prime}}-{\bf f}_{a a^{\prime}}}{2}-\gamma^5
\frac{\delta_{a a^{\prime}}+{\bf f}_{a a^{\prime}}}{2}]\psi_{t,a^{\prime}} +
(c.c.))  d^4 x\nonumber\\ \label{Sf22SM___}
\end{eqnarray}

Here we have composed new spinors $\psi^a_t = \sqrt{{[{\rm Re}\zeta]-[{\rm Im} \chi]}}\left(\begin{array}{c}\phi^a_-\\
\psi^a_+\end{array}\right)$ and $\psi^a_s = {\gamma^5}\sqrt{{[{\rm Re}\zeta]+[{\rm Im} \chi]}}\left(\begin{array}{c}\psi^a_-\\
\phi^a_+\end{array}\right)$.

Next, we neglect SM gauge fields that are to be considered as perturbations. We
also introduce the auxiliary fields as in the ENJL approach:
\begin{eqnarray}
S_f & = & \int \{i\bar{\psi}^a_s \gamma^{\mu} D_{\mu} \psi^a_s -
\frac{3\alpha_-}{16M_T^2}(\bar{\psi}^a_s\gamma^i \gamma^5
\psi^a_s)(\bar{\psi}^b_s\gamma_i \gamma^5  \psi^b_s)\} d^4 x \nonumber\\&&+
\int \{i\bar{\psi}^a_t \gamma^{\mu} D_{\mu} \psi^a_t \} d^4 x
\nonumber\\&&+\int\{ -(\bar{\psi}^a_{t,L}H^+_{ab} \psi^b_{t,R} + (h.c.)) -
\frac{4M_T^2}{3\alpha_+} {\rm Tr}\, H^+H\}d^4x \nonumber\\&&
+\int\{(\bar{\psi}^a_{t,L}\gamma^i L^{ab}_i\psi^b_{t,L}) -
\frac{4M_T^2}{3\alpha_+}{\rm Tr}\,L^iL_i +(L\leftarrow \rightarrow R)\} d^4
x\nonumber\\&& - \frac{m_0}{\sqrt{[{\rm Re}\zeta]^2-[{\rm Im}
\chi]^2}}\int(\bar{\psi}_{s,a} [\frac{\delta_{a a^{\prime}}-{\bf f}_{a
a^{\prime}}}{2}-\gamma^5 \frac{\delta_{a a^{\prime}}+{\bf f}_{a
a^{\prime}}}{2}]\psi_{t,a^{\prime}} + (c.c.))  d^4 x\nonumber\\
\label{Sf22SMNJL}
\end{eqnarray}

 Integration over technifermions leads to appearance of the effective potential for $H$ that has its minimum at
 $H = m_t {\bf 1}$. So, $H = m_t {\bf 1} + h$, where vacuum value of $h$ is zero. Thus we get:
 \begin{eqnarray}
&&S_f  =  \int  \{i\bar{\psi}^a_s \gamma^{\mu} D_{\mu} \psi^a_s -
\frac{3\alpha_-}{16M_T^2}(\bar{\psi}^a_s\gamma^i \gamma^5
\psi^a_s)(\bar{\psi}^b_s\gamma_i \gamma^5  \psi^b_s)\} d^4 x+ S_{eff}[L,R,H]
\nonumber\\&&-\frac{m^2_0}{{[{\rm Re}\zeta]^2-[{\rm Im}
\chi]^2}}\int\{ {\left(\begin{array}{c}\psi_{s,L}\\
-{\bf f} \psi_{s,R}\end{array}\right)}^+\gamma^0   [i \gamma^{\mu} D_{\mu}- m_t
{\bf 1}-h] ^{-1}{\left(\begin{array}{c}\psi_{s,L}\\
-{\bf f} \psi_{s,R}\end{array}\right)} \}d^4x \nonumber\\\label{Sf22SMNJL_}
\end{eqnarray}
where $D_{\mu} = (\partial_{\mu} -i \frac{1+\gamma_5}{2}L_{\mu}-i
\frac{1-\gamma_5}{2}R_{\mu})$. Here we denote $S_{eff}= -i{\rm Sp}\,{\rm Log}[i
\gamma^{\mu} D_{\mu}- m_t {\bf 1}-h] $.

 Now our supposition is that $m_t >> m_0$. Next, at the energies much less than $M_T$ we can omit the four fermion terms for $\psi_s$.
 We also neglect fluctuations of $h$, $L$, and $R$  around their zero vacuum values and arrive at:

\begin{eqnarray}
&&S_f  =  \int \bar{\psi}_s (i\gamma^{\mu} \partial_{\mu}-\frac{m^2_0}{{[{\rm
Re}\zeta]^2-[{\rm Im} \chi]^2}} {\bf f} [ m_t ]^{-1})\psi_sd^4x
\label{Sf22SMNJLF}
\end{eqnarray}

 As a result the mass term for
$\psi_s$ appears with the mass matrix

\begin{equation}
m_s = \frac{m^2_0}{{[{\rm Re}\zeta]^2-[{\rm Im} \chi]^2}}\frac{{\bf f}}{m_t}
\end{equation}

It is worth mentioning that in order to have $m_s$ positive defined we need
$[{\rm Re}\zeta]^2>[{\rm Im} \chi]^2$ provided that $\bf f$ is also positive
defined. When $[{\rm Re}\zeta]^2<[{\rm Im} \chi]^2$ we can compose $\psi_s$ as
follows: $\psi^a_s = \sqrt{{[{\rm Re}\zeta]+[{\rm Im} \chi]}}\left(\begin{array}{c}\psi^a_-\\
\phi^a_+\end{array}\right)$ and arrive at $m_s = \frac{m^2_0}{{[{\rm Im}
\chi]^2-[{\rm Re}\zeta]^2}}\frac{{\bf f}}{m_t}$.

\section{Conclusions}

In this paper we considered fermions coupled in a nonminimal way to Poincare
gravity. The gravity action contains the Holst action with the mass parameter
at a Tev scale. In addition the fermion action itself breaks Parity. That's why
the  left - right asymmetry appears in the effective four - fermion
interactions. We arrange all SM fermions in the left - handed components of the
Dirac spinors while the right - handed components are reserved for
technifermions. Via an appropriate choice of couplings the four - fermion terms
that contain SM fermions can be made repulsive while the four - fermion terms
that contain technifermions can be made attractive. We also need that the four
fermion interaction that contains both SM fermions and technifermions is
negligible (\ref{req}). Therefore, the technifermions are condensed and cause
the appearance of $W$ and $Z$ - boson masses. We need
$\frac{M_T}{\sqrt{\alpha_+}} < \Lambda_{\chi}$, where $\Lambda_{\chi}$ is the
scale, at which the dynamical Poincare gravity theory appears. We expect ${M_T}
\sim 1$ Tev. At the same time the scale $\Lambda_{\chi}$ is expected to be
between $1$ Tev and $10$ Tev.

An obvious difficulty of our approach is that we need the effective coupling
constants to satisfy  condition (\ref{req}). That's why the detailed analysis
of the renormalization group trajectories is needed in order to investigate the
necessary domain of the theory. We also need to know what does it mean:
$|\alpha_{+-}| << \alpha_+$. For example, is this sufficient or not to have
$|\alpha_{+-}| \sim 0.1 \, \alpha_+$, may become clear only after the detailed
analysis of the NJL model is performed. Namely, we need to investigate the NJL
model that includes  $\psi_s$ and $\psi_t$ with the four fermion term that
includes both $\psi_t$ and $\psi_s$.

In order to provide appearance of masses for the SM fermions we  add the mass
term for the Dirac spinors that contain SM fermions as their left-handed
components. This term is considered as a perturbation over the four - fermion
interactions caused by torsion. As a result the mass term for the SM fermions
appears.

There is the important question about the scale of ${M_T}$, $\Lambda_{\chi}$,
and the mass parameter entering (\ref{Sf2SM__}) that gives rise to SM mass
matrix. Actually, if one assumes that quantum gravity theory (that is the
dynamical theory of metric) exists at the energies of the order of Planck mass
$m_p$, such mass parameters might be generated dynamically and, therefore,
receive values at a $m_p$ scale. Therefore  we must suppose that there exists a
mechanism within the $m_p$ scale theory that forbids dynamical generation of
torsion mass as well as $m_0$ from (\ref{Sf2SM__}). Actually, we may suppose
that there is no quantum theory of Riemannian geometry at all. Then the
dynamical torsion theory may be thought of as a gauge theory of Lorentz group
that is defined in Minkowsky space \cite{Minkowsky, Minkowsky_}. This theory
may have a scale slightly above $1$ TeV. In this approach there is no problem
with the scale $m_p$ at all. In such a scheme classical gravity may appear, for
example, as an entropic force \cite{Entropy_force}.

This work was partly supported by RFBR grants  09-02-00338, 08-02-00661, by
Grant for leading scientific schools 6260.2010.2.

\clearpage

\end{document}